\documentclass[twocolumn,showpacs,preprintnumbers,amsmath,amssymb]{revtex4}
\usepackage{graphicx}% Include figure files
\usepackage{dcolumn}% Align table columns on decimal point
\usepackage{bm}% bold math

\begin{document}

\title{Invariant Operators vs Heisenberg Operators
for Time-Dependent Generalized Oscillators}

\author{Sang Pyo Kim}\email{sangkim@kunsan.ac.kr}

\affiliation{Department of Physics, Kunsan National University,
Kunsan 573-701, Korea}

\affiliation{Asia Pacific Center for Theoretical Physics, Pohang
790-784, Korea}

\date{\today}
\begin{abstract}
We investigate the relation between the invariant operators
satisfying the quantum Liouville-von Neumann and the Heisenberg
operators satisfying the Heisenberg equation. For time-dependent
generalized oscillators we find the invariant operators, known as
the Ermakov-Lewis invariants, in terms of a complex classical
solution, from which the evolution operator is derived, and obtain
the Heisenberg position and momentum operators. Physical
quantities such as correlation functions are calculated using both
the invariant operators and Heisenberg operators.
\end{abstract}

\maketitle

\section{Introduction}

In the last several decades there have been many attempts to
develop formalisms for time-dependent quantum systems and apply
them to physical systems, in particular, oscillators with
time-dependent mass and frequency. For a time-dependent oscillator
Lewis found an invariant operator, which satisfies the quantum
Liouville-von Neumann (LvN) equation, and whose eigenfunctions up
to time-dependent phase factors satisfy the time-dependent
Schr\"{o}dinger equation \cite{lewis}. In classical theory Ermakov
had already found such an invariant long before the advent of
quantum theory \cite{ermakov}.  The Ermakov-Lewis invariant
operators since then have been employed to develop the formalism
for time-dependent oscillators and applied to many physical
systems such as quantum optics. The original Ermakov-Lewis
invariants are quadratic in position and momentum operators
\cite{quad-inv} and the linear invariants are also introduced
\cite{lin-inv,kim}.

The invariant operators method is a useful and convenient tool to
find various kinds of exact quantum states in terms of the
solutions of the auxiliary equation and has been widely applied to
physical systems. The relation between the Ermakov-Lewis invariant
operators and the Heisenberg operators in terms of the evolution
operator was noticed by Dodonov and Man'ko \cite{dodonov}. Also
the Heisenberg picture was used applied to time-dependent
oscillators \cite{heis1,heis2}. However, the relation between the
quantum LvN equation for the invariant operators and the
Heisenberg equation for the Heisenberg operators has not been
seriously investigated.

The purpose of this paper is to clarify the role of the quantum
LvN equation used in the invariant operators method and compare it
with the Heisenberg picture. The Schr\"{o}dinger and Heisenberg
pictures are the two popular descriptions of quantum theory. In
the Schr\"{o}dinger picture, quantum states evolve in time but
operators do not change, whereas in the Heisenberg picture,
operators evolve in time but quantum states do not change
\cite{sakurai}. The quantum theory of systems are equivalently
described in the both pictures. It is shown that the quantum LvN
equation provides another description of quantum theory
independently of the Heisenberg picture. In terms of the evolution
operator $\hat{U}(t)$, the Heisenberg operators evolve according
to $\hat{O}_{\rm H} (t) = \hat{U}^{\dagger}(t) \hat{O} \hat{U}
(t)$ and satisfy the Heisenberg equation, whereas the invariant
operators evolve according to $\hat{O}_{\rm L} (t) = \hat{U}(t)
\hat{O} \hat{U}^{\dagger}(t)$ and satisfy the LvN equation. To
illustrate this point, we apply the quantum theory based on the
LvN equation to the time-dependent generalized oscillators.
Further, we use the evolution operator from the invariant
operators to find the Heisenberg operators in terms of the
classical solution.

The organization of the paper is as follows. In Sec. II we discuss
the Schr\"{o}dinger and Heisenberg pictures. We also discuss the
difference and relation between the Heisenberg equation and the
LvN equation. In Sec. III, using a time-independent generalized
oscillator, we illustrate the difference between the invariant
operators and the Heisenberg operators. In Sec. IV we apply the
invariant operators and the Heisenberg operators to time-dependent
oscillators. The linear invariant operators are found directly
from the LvN equation in terms of a classical solution. The
evolution operator derived from the invariant operator is used to
find the Heisenberg operators.

\section{Invariant Operators vs Heisenberg Operators}

For stationary (time-independent) quantum systems two pictures
have been used: the Schr\"{o}dinger and Heisenberg pictures. In
this paper we consider both time-independent and time-dependent
systems simultaneously. The Hamiltonian operator for a
time-dependent system has time-dependent coefficients of each
Schr\"{o}dinger operator $\hat{O}_{k}$:
\begin{equation}
\hat{H} (t) = \sum_{k} h_k (t) \hat{O}_{k}.
\end{equation}
In this sense the time-dependent Hamiltonian is called a
Schr\"{o}dinger operator, though it depends on time explicitly.

In the Schr\"{o}dinger picture, the quantum states evolve
according to the Schr\"{o}dinger equation with the Hamiltonian
$\hat{H}$
\begin{equation}
i \hbar \frac{\partial}{\partial t} \vert \Psi, t \rangle =
\hat{H} (t) \vert \Psi, t \rangle.
\end{equation}
All the information of the system is contained in the state $\vert
\Psi, t \rangle$. The physically measurable quantity corresponding
to an observable $\hat{O}$ is given by the expectation value
\begin{equation}
\langle \hat{O} \rangle_{\rm S} (t) = \langle \Psi, t \vert
\hat{O} \vert \Psi, t \rangle
\end{equation}
We introduce the evolution operator satisfying
\begin{equation}
i \hbar \frac{\partial}{\partial t} \hat{U} (t) = \hat{H} (t)
\hat{U} (t). \label{ev eq}
\end{equation}
The solution is formally written as the time-ordered integral
\begin{equation}
\hat{U} (t) = {\rm T} \exp  \Bigl(- \frac{i}{\hbar} \int^t \hat{H}
(t') dt' \Bigr). \label{ev op}
\end{equation}
Note that for a time-dependent system, $\hat{H} (t) $ and $
\hat{H} (t')$ do not commute in general for $ t \neq t'$. For a
stationary (time-independent) system with $\hat{H}_0$, the
evolution operator is simply given by $\hat{U} (t) = e^{- i
\hat{H}_0 t/\hbar}$. The state vector is then written as
\begin{equation}
\vert \Psi, t \rangle = \hat{U} (t) \vert \Psi \rangle, \label{sch
st}
\end{equation}
where $\vert \Psi \rangle$ is any state independent of time.

On the other hand, in the Heisenberg picture, quantum states do
not change in time but operators evolve as
\begin{equation}
\hat{O}_{\rm H} (t) = \hat{U}^{\dagger} (t) \hat{O} \hat{U} (t).
\label{heis op}
\end{equation}
The Heisenberg operators $\hat{O}_{\rm H}(t)$ carrying all the
information of the system satisfy the Heisenberg equation
\begin{equation}
i \hbar \frac{\partial}{\partial t} \hat{O}_{\rm H} (t) + [
\hat{H}_{\rm H} (t), \hat{O}_{\rm H} (t)] = 0, \label{heis eq}
\end{equation}
where $\hat{H}_{\rm H} (t)$ is the Heisenberg operator
\begin{equation}
\hat{H}_{\rm H} (t) = \sum_{k} h_k (t) \hat{O}_{k{\rm H}}(t).
\end{equation}
For the stationary system $\hat{U} (t)$ commutes with $H_0$, so
$\hat{H}_{\rm H} = \hat{H}_0$. The expectation value of the
observable $\hat{O}$ in the Heisenberg picture now takes the form
\begin{equation}
\langle \hat{O} \rangle_{\rm H} (t) = \langle \Psi \vert
\hat{O}_{\rm H} (t) \vert \Psi \rangle = \langle \hat{O}
\rangle_{\rm S} (t).
\end{equation}

We now introduce another interesting operator, the so-called
invariant or the LvN operator,
\begin{equation}
\hat{O}_{\rm L} (t) = \hat{U} (t) \hat{O} \hat{U}^{\dagger} (t).
\label{lvn op}
\end{equation}
The invariant operator $\hat{O}_{\rm L}(t)$ is the backward
evolution of the Schr\"{o}dinger operator $\hat{O}$, whereas the
Heisenberg operator $\hat{O}_{\rm H}(t)$ is the forward evolution
\cite{balian}. It follows that the invariant operators satisfy the
quantum LvN equation
\begin{equation}
i \hbar \frac{\partial}{\partial t} \hat{O}_{\rm L} (t) + [
\hat{O}_{\rm L} (t), \hat{H} (t)] = 0. \label{lvn eq}
\end{equation}
Note that the Hamiltonian $\hat{H} (t)$ appeared in Eq. (\ref{lvn
eq}) is a Schr\"{o}dinger operator, whereas the Heisenberg
Hamiltonian operator appears in Eq. (\ref{heis eq}).

Let us now suppose that the LvN equation (\ref{lvn eq}) be
directly solved by some techniques. For time-dependent
oscillators, these operators are explicitly given in terms of the
solution of an auxiliary equation and widely known as the
Ermakov-Lewis invariants. In terms of the eigenstates of an
invariant operator $\hat{O}_{\rm L} (t)$,
\begin{equation}
\hat{O}_{\rm L} (t) \vert \lambda, t \rangle = \lambda \vert
\lambda, t \rangle,
\end{equation}
the exact quantum state of the Schr\"{o}dinger equation is given
by
\begin{equation}
\vert \Psi, t \rangle = \exp \Bigl( \langle  \lambda, t \vert - (
\frac{i}{\hbar} \hat{H} (t) + \frac{\partial}{\partial t}) \vert
\lambda, t  \rangle \Bigr) \vert \lambda, t \rangle.
\end{equation}
It follows from Eqs. (\ref{lvn op}) and (\ref{sch st}) that
\begin{equation}
\hat{O} \vert \lambda \rangle = \vert \lambda \rangle.
\end{equation}
Thus invariant operators carry all the information and provide a
Hilbert space of quantum states without the direct knowledge of
the evolution operator. The expectation value of the observable
takes the same value as the Schr\"{o}dinger picture
\begin{equation}
\langle \hat{O} \rangle_{\rm L} (t) = \langle \Psi, t \vert
\hat{O} \vert \Psi, t \rangle = \langle \hat{O} \rangle_{\rm S}
(t).
\end{equation}

In conclusion, it is shown that the quantum LvN equation (\ref{lvn
eq}) self-consistently provides another description of quantum
theory because all the invariant operators $\hat{O}_{\rm L}$
satisfying Eq. (\ref{lvn eq}) carry all information of the system.
Therefore, the invariant operators method may be called the  LvN
picture just as quantum theory based  Eq. (\ref{heis eq}) is
called the Heisenberg picture, in both of which time-dependent
operators carry all quantum information.

\section{Time-Independent Generalized Oscillator}

As a simple and illustrative model, we consider the
time-independent generalized oscillator
\begin{equation}
\hat{H}_0 = \frac{X_0}{2} \hat{p}^2 + \frac{Y_0}{2}
(\hat{p}\hat{q} + \hat{q}\hat{p}) + \frac{Z_0}{2} \hat{q}^2,
\label{in osc}
\end{equation}
where it is assumed
\begin{equation}
\omega_0^2 = X_0 Z_0 - Y_0^2 \geq 0.
\end{equation}
We introduce the annihilation (lowering) and creation (raising)
operators
\begin{eqnarray}
\hat{a} &=& \sqrt{\frac{X_0}{2 \hbar \omega_0}} \Biggl[i \hat{p} +
\frac{1}{X_0} (\omega_0 +i Y_0) \hat{q} \Biggr], \nonumber\\
\hat{a}^{\dagger} &=& \sqrt{\frac{X_0}{2 \hbar \omega_0}} \Biggl[-
i  \hat{p} + \frac{1}{X_0} (\omega_0 -i Y_0) \hat{q} \Biggr].
\label{in ca}
\end{eqnarray}
The annihilation and creation operators satisfy the usual
commutation relation $[ \hat{a}, \hat{a}^{\dagger} ] = 1$.
Inverting Eq. (\ref{in ca}) for the position and momentum
operators, we obtain
\begin{eqnarray}
\hat{q} &=& \sqrt{\frac{\hbar X_0}{2 \omega_0}} (\hat{a} +
\hat{a}^{\dagger}), \nonumber\\
\hat{p} &=& \sqrt{\frac{\hbar}{2 \omega_0 X_0}} [ -(i \omega_0
+Y_0) \hat{a} + (i \omega_0 - Y_0) \hat{a}^{\dagger} ]. \label{in
pm}
\end{eqnarray}
Then the Hamiltonian takes the form
\begin{equation}
\hat{H}_0 = \hbar \omega_0 \Biggl(\hat{a}^{\dagger} \hat{a} +
\frac{1}{2} \Biggr).
\end{equation}

Now, using the evolution operator
\begin{equation}
\hat{U} (t) = e^{ - i \omega_0 ( \hat{a}^{\dagger} \hat{a} +
1/2)},
\end{equation}
we find the Heisenberg operators
\begin{equation}
\hat{a}_{\rm H} (t) = e^{- i \omega_0 t} \hat{a}, \quad
\hat{a}_{\rm H}^{\dagger} (t) = e^{i \omega_0 t}
\hat{a}^{\dagger}, \label{in heis}
\end{equation}
and the invariant operators
\begin{equation}
\hat{a}_{\rm L} (t) = e^{i \omega_0 t} \hat{a}, \quad \hat{a}_{\rm
L}^{\dagger} (t) = e^{- i \omega_0 t} \hat{a}^{\dagger}. \label{in
lvn}
\end{equation}
The Heisenberg operators (\ref{in heis}) satisfy the Heisenberg
equation, whereas the invariant operators (\ref{in lvn}) satisfy
the quantum LvN equation. The invariant operator $\hat{a}_{\rm
L}(t)$ has the positive frequency in contrast with the negative
frequency of the Heisenberg operator $\hat{a}_{\rm H}(t)$.

The Heisenberg position and momentum operators are obtained from
Eq. (\ref{in pm}) through the unitary transformation (\ref{heis
op}) as
\begin{eqnarray}
\hat{q}_{\rm H} (t) &=& \sqrt{\frac{\hbar X_0}{2 \omega_0}}
[\hat{a}_{\rm H} +
\hat{a}^{\dagger}_{\rm H}], \nonumber\\
\hat{p}_{\rm H} (t) &=& \sqrt{\frac{\hbar}{2 \omega_0 X_0}} [ -(i
\omega_0 +Y_0) \hat{a}_{\rm H} + (i \omega_0 - Y_0)
\hat{a}^{\dagger}_{\rm H} ]. \label{in heis pm}
\end{eqnarray}
On the other hand, the Schr\"{o}dinger position and momentum
operators  in terms of the invariant annihilation and creation
operators are given by
\begin{eqnarray}
\hat{q} &=&  \sqrt{\hbar} [u \hat{a}_{\rm L} +
u^* \hat{a}^{\dagger}_{\rm L}], \nonumber\\
\hat{p} &=& \frac{\sqrt{\hbar}}{X_0} [ (\dot{u} - Y_0 u)
\hat{a}_{\rm L} + (\dot{u}^* - Y_0 u^*) \hat{a}^{\dagger}_{\rm L}
], \label{in lvn pm}
\end{eqnarray}
where $u(t)$ is the complex solution to the classical equation of
motion
\begin{equation}
u (t) = \sqrt{\frac{X_0}{ 2 \omega_0}} e^{- i \omega_0 t}.
\end{equation}

\section{Time-Dependent Generalized Oscillator}

We now consider a time-dependent generalized oscillator described
by the Hamiltonian \cite{cervero,kim-page}
\begin{equation}
\hat{H} (t) = \frac{X(t)}{2} \hat{p}^2 + \frac{Y(t)}{2}
(\hat{p}\hat{q} + \hat{q}\hat{p}) + \frac{Z(t)}{2} \hat{q}^2,
\label{osc}
\end{equation}
where $X, Y$ and $Z$ explicitly depend on time. The classical
equation of motion corresponding to the Hamiltonian operator
(\ref{osc}) is given by
\begin{equation}
\frac{d}{dt} \Biggl(\frac{\dot{u}}{X} \Biggr) + \Biggl[ XZ - Y^2 +
\frac{\dot{X} Y - X \dot{Y} }{X}\Biggr] \Biggl(\frac{u}{X} \Biggr)
= 0. \label{cl eq}
\end{equation}

In the LvN picture, we look for the operators satisfying the LvN
equation (\ref{lvn eq}). Lewis and Riesenfeld found a quaratic
invariant \cite{lewis}. Following Ref. \cite{kim-page} we
introduce a pair of linear invariant operators
\begin{eqnarray}
\hat{a}_{\rm L} (t)  &=& \frac{i}{\sqrt{\hbar}} \Biggl[u^* \hat{p}
- \frac{1}{X} ( \dot{u}^* - Y u^*) \hat{q} \Biggr], \nonumber\\
\hat{a}^{\dagger}_{\rm L} (t) &=& - \frac{i}{\sqrt{\hbar}}
\Biggl[u \hat{p} - \frac{1}{X} ( \dot{u} - Y u) \hat{q} \Biggr],
\label{inv op}
\end{eqnarray}
where $u$ is a complex solution to the classical equation of
motion (\ref{cl eq}). The Wronskian condition
\begin{equation}
{\rm Wr} \{u^*, u\} = \frac{1}{X} (u \dot{u}^* - u^* \dot{u}) = i,
\label{wr}
\end{equation}
makes the invariant operators satisfy the standard commutation
relation at equal time
\begin{equation}
[\hat{a}_{\rm L} (t), \hat{a}^{\dagger}_{\rm L} (t) ] = 1.
\end{equation}
The Schr\"{o}inger position and momentum operators can be
expressed in terms of the invariant annihilation and creation
operators as
\begin{eqnarray}
\hat{q} &=& \sqrt{\hbar} (u \hat{a}_{\rm L} + u^*
\hat{a}^{\dagger}_{\rm L}),
\nonumber\\
\hat{p} &=& \frac{\sqrt{\hbar}}{X} [(\dot{u} - Y u) \hat{a}_{\rm
L} + (\dot{u}^* - Y u^*) \hat{a}^{\dagger}_{\rm L}]. \label{lvn
pm}
\end{eqnarray}
Note that Eq. (\ref{in lvn pm}) is the constant parameters limit
of Eq. (\ref{lvn pm}).

The Fock space of exact quantum states for the Schr\"{o}dinger
equation consists of the number state $\vert n, t \rangle$ of the
number operator $\hat{N}_{\rm L} (t) = \hat{a}^{\dagger}_{\rm L}
(t) \hat{a}_{\rm L} (t)$, another invariant operator:
\begin{equation}
\hat{N}_{\rm L} (t) \vert n, t \rangle = n \vert n, t \rangle.
\end{equation}
The quadratic Ermakov-Lewis invariants \cite{lewis,quad-inv} are,
up to overall constants, the number operator $\hat{N} (t)$, whose
auxiliary field $\rho$ is the amplitude of the complex solution
$u$ \cite{kim-page}. In contrast with the Schr\"{o}dinger and
Heisenber pictures, both the operators and states of the LvN
picture depend on time but their eigenvalues (quantum numbers) do
not change. For instance, the correlation functions with respect
to the number state are given by
\begin{eqnarray}
\langle n, t \vert \hat{q}^2 \vert n, t \rangle &=& \hbar u^* u
(2n + 1), \nonumber\\
\langle n, t \vert \hat{p}^2 \vert n, t \rangle &=&
\frac{\hbar}{X^2}
(\dot{u}^* - Y u^*)(\dot{u} - Yu) (2n + 1), \nonumber\\
\langle n, t \vert \frac{1}{2}(\hat{p} \hat{q} + \hat{q} \hat{p})
\vert n, t \rangle &=& \frac{\hbar}{2X} [(\dot{u}^* - Y u^*) u +
(\dot{u} - Y u) u^*] \nonumber\\ && \times (2n + 1). \label{osc
lvn eq}
\end{eqnarray}

The invariant operators (\ref{inv op}) are related with the
annihilation and creation operators (\ref{in ca}) through the
Bogoliubov transformation
\begin{eqnarray}
\hat{a}_{\rm L} (t) &=& \alpha \hat{a} + \beta \hat{a}^{\dagger},
\nonumber\\ \hat{a}^{\dagger}_{\rm L} (t) &=& \alpha^*
\hat{a}^{\dagger} + \beta^* \hat{a}, \label{lvn bog}
\end{eqnarray}
where
\begin{eqnarray}
\alpha &=& - i \sqrt{\frac{X_0}{2 \omega_0}} \Biggl[ \frac{1}{X_0}
(i \omega_0 + Y_0)u^* + \frac{1}{X} (\dot{u}^* - Y u^*) \Biggr], \nonumber\\
\beta &=& - i \sqrt{\frac{X_0}{2 \omega_0}} \Biggl[ \frac{1}{X_0}
(- i \omega_0 + Y_0)u^* + \frac{1}{X} (\dot{u}^* - Y u^*) \Biggr].
\end{eqnarray}
The evolution operator (\ref{ev op}) can be written as
\begin{equation}
\hat{U} (t) = e^{ - i \vartheta (t) \hat{a}^{\dagger} \hat{a}}
\hat{S} (t) \label{osc ev}
\end{equation}
where
\begin{equation}
\hat{S} (t) = e^{\frac{1}{2} [z (t) \hat{a}^2 - z^* (t)
\hat{a}^{\dagger 2}]},
\end{equation}
is the squeeze operator. The squeeze operator leads to the
Bogoliubov transformation \cite{stoler}
\begin{eqnarray}
\hat{S} (t) \hat{a} \hat{S}^{\dagger} (t) &=& \cosh r~ \hat{a} +
e^{ - i
\varphi} \sinh r ~\hat{a}^{\dagger}, \nonumber\\
\hat{S} (t) \hat{a}^{\dagger} \hat{S}^{\dagger} (t) &=& \cosh r
~\hat{a}^{\dagger} + e^{ i \varphi} \sinh r ~\hat{a},
\end{eqnarray}
where $z = r e^{i \varphi}$. From another form (\ref{lvn op}) of
the invariant
\begin{equation}
\hat{a}_{\rm L} (t) = \hat{U} (t) \hat{a} \hat{U}^{\dagger} (t),
\end{equation}
we find the phase and squeeze parameters
\begin{equation}
\alpha = e^{ i \vartheta} \cosh r, \quad \beta = e^{- i \vartheta}
e^{-i \varphi} \sinh r.
\end{equation}
For the time-dependent oscillator with $Y = 0$, the evolution
operator in the Wei-Norman form \cite{wei}
\begin{equation}
\hat{U} (t) = e^{ i w (t) (\hat{p} \hat{q} + \hat{q}\hat{p})/2}
e^{i w_+ (t) \hat{q}^2} e^{- i w_- \hat{p}^2}
\end{equation}
has been used in Ref. \cite{ev op}, where $w$ and $w_{\pm}$ are
found by directly solving Eq. (\ref{ev eq}). It was found there
that $e^{- w}$ and $w_- e^{-w}$ satisfy the classical equation
(\ref{cl eq}).

Now, using the evolution operator (\ref{osc ev}),  the
annihilation and creation operators in the Heisenberg picture are
found to be
\begin{eqnarray}
\hat{a}_{\rm H} (t) &=& \hat{U}^{\dagger} (t) \hat{a} \hat{U} (t)
=
\alpha^* \hat{a} - \beta \hat{a}^{\dagger}, \nonumber\\
\hat{a}_{\rm H}^{\dagger} (t) &=& \hat{U}^{\dagger} (t)
\hat{a}^{\dagger} \hat{U} (t) = \alpha \hat{a}^{\dagger} - \beta^*
\hat{a}.
\end{eqnarray}
Similarly, from Eq. (\ref{in pm}) follows the Heisenberg position
operator
\begin{equation}
\hat{q}_{\rm H} (t) = \hat{U}^{\dagger} (t) \hat{q} \hat{U} (t) =
\sqrt{\hbar} (u^* (t) \hat{a} + u (t) \hat{a}^{\dagger}).
\label{heis pos}
\end{equation}
The Heisenberg momentum operator takes the form
\begin{equation}
\hat{p}_{\rm H} (t) = \frac{1}{X} ( \dot{\hat{q}}_{\rm H} - Y
\hat{q}_{\rm H} ) = \frac{\sqrt{\hbar}}{X} [( \dot{u}^* - Y u^*)
\hat{a} + (\dot{u} - Y u) \hat{a}^{\dagger}]. \label{heis mom}
\end{equation}
Note that another Heisenberg operator $\hat{U}^{\dagger} \hat{p}
\hat{U}$ is not conjugate to $\hat{q}_{\rm H}$ except for the
time-independent case. In the Heisenberg picture physical
quantities are obtained by taking the expectations of Heisenberg
operators with respect to time-independent state relevant to the
physical system under study. The correlation functions with
respect to the number state $\vert n \rangle $ of
$\hat{a}^{\dagger} \hat{a}$ in Sec. II are given by
\begin{eqnarray}
\langle n \vert \hat{q}^2_{\rm H} \vert n \rangle &=& \hbar u^* u
(2n + 1), \nonumber\\
\langle n \vert \hat{p}^2_{\rm H} \vert n \rangle &=&
\frac{\hbar}{X^2}
(\dot{u}^* - Y u^*)(\dot{u} - Yu) (2n + 1), \nonumber\\
\langle n \vert \frac{1}{2}(\hat{p}_{\rm H} \hat{q}_{\rm H} +
\hat{q}_{\rm H} \hat{p}_{\rm H}) \vert n \rangle &=&
\frac{\hbar}{2X} [(\dot{u}^* - Y u^*) u + (\dot{u} - Y u) u^*]
\nonumber\\ && \times (2n + 1). \label{osc heis eq}
\end{eqnarray}

A few comments are in order. First, as $u$ and $u^*$ satisfy Eq.
(\ref{cl eq}), the Heisenberg position operator (\ref{heis pos})
satisfies the Heisenberg equation (\ref{heis eq}) for the
time-dependent oscillator (\ref{osc}):
\begin{equation}
\frac{d}{dt} \Biggl(\frac{\dot{\hat{q}}_{\rm H}}{X} \Biggr) +
\Biggl[ XZ - Y^2 + \frac{\dot{X}Y - X \dot{Y}}{X}\Biggr]
\Biggl(\frac{\hat{q}_{\rm H}}{X} \Biggr) = 0. \label{heis eq2}
\end{equation}
In fact, the general solution to Eq. (\ref{heis eq2}) is
\begin{equation}
\hat{q}_{\rm H} (t) = v_1 (t) \hat{q}_1 + v_2 (t) \hat{q}_2,
\end{equation}
where $v_1$ and $v_2$ are the solutions to Eq. (\ref{cl eq}) and
$\hat{q}_1$ and $\hat{q}_2$ are two independent operators whose
commutator does not vanish. The operator (\ref{heis pos}) is the
special case with $v_1 = \sqrt{\hbar} u^*$, $v_2 = \sqrt{\hbar} u$
and $\hat{q}_1 = \hat{a}$, $\hat{q}_2 = \hat{a}^{\dagger}$. This
choice makes the standard commutation relation satisfied at equal
time
\begin{equation}
[ \hat{q}_{\rm H} (t), \hat{p}_{\rm H} (t) ] = i \hbar.
\end{equation}
Second, by comparing Eqs. (\ref{lvn pm}) and Eqs. (\ref{heis pos})
and (\ref{heis mom}), we see that the invariant operators have the
negative (positive) frequency for the creation (annihilation)
operator, which is a consequence of the backward evolution of the
invariant operator (\ref{inv op}). Third, the correlation
functions (\ref{osc lvn eq}) in the LvN picture are the same as
those (\ref{osc heis eq}) in the Heisenberg picture.

\section{Conclusion}

In this paper we show that the invariant operators method based on
the quantum Liouville-von Neumann equation provides another
description of quantum theory, a counterpart of the Heisenberg
picture. The invariant operators method, which we prefer to call
the invariant picture, shares not only the property of the
Schr\"{o}dinger picture that quantum states are time-dependent,
but also the property of the Heisenberg picture that operators are
time-dependent, backwardly evolving in time. In terms of the
evolution operator $\hat{U}(t)$, the Heisenberg operators evolve
according to $\hat{O}_{\rm H} (t) = \hat{U}^{\dagger} (t) \hat{O}
\hat{U}(t)$ and satisfy the Heisenberg equation, whereas the
invariant operators to $\hat{O}_{\rm L} (t) = \hat{U} (t) \hat{O}
\hat{U}^{\dagger}(t)$ and satisfy the quantum Liouville-von
Neumann equation. As $\hat{O}_{\rm L} (t)$ carries all information
of the system, the quantum Liouville-von Neumann equation provides
another picture in addition to the Schr\"{o}dinger and the
Heisenberg pictures.

To illustrate the invariant picture we apply it to
time-independent and time-dependent generalized oscillators. The
time-independent oscillator shows manifestly the difference
between the Heisenberg and the invariant pictures. The invariant
operators have the opposite signs of frequencies to the Heisenberg
operators. The invariant picture is particularly useful to handle
time-dependent quantum systems. The Hilbert space of exact quantum
states of the time-dependent Schr\"{o}dinger equation consists of
the eigenstates of an invariant operator. For time-dependent
generalized oscillators we find two linear invariant operators and
use them to derive the evolution operator. The evolution operator
is then used to find the Heisenberg position and momentum
operators. It is shown that the invariant picture and the
Heisenberg picture provide the same physical results.

\acknowledgements

This work was supported by the Korea Research Foundation under
Grant No. KRF-2002-041-C00053.

\end{document}